\pdfoutput=1
\documentclass[11pt,twoside]{article}
\usepackage{asp2010}

\usepackage{graphicx}

\resetcounters

\begin{document}

\title{The Build-up to Eruptive Solar Events Viewed as the Development of Chiral Systems}

\author{Sara~F.~Martin,$^1$ Olga~Panasenco,$^1$ Mitchell~A.~Berger,$^2$
Oddbjorn~Engvold,$^3$ Yong~Lin,$^3$ Alexei~A.~Pevtsov,$^4$ and Nandita~Srivastava$^5$
\affil{$^1$Helio Research, 5212 Maryland Ave., La Crescenta, CA, 91214, USA\\
$^2$University of Exeter, SECAM, Exeter, EX4 4QE, UK\\
$^3$Institute for Theoretical Astrophysics, University of Oslo, P.O. Box 1029, Blindern, N-0315 Oslo 3, Norway\\
$^4$National Solar Observatory, Sunspot, NM, 88349, USA\\
$^5$Udaipur Solar Observatory, Physical Research Laboratory, P.B. 198, Badi Road, Dewali, Udaipur 313001, India}
}
\begin{abstract}
When we examine the \emph{chirality} or observed handedness of the chromospheric and coronal structures involved in the long-term build-up to eruptive events, we find that they evolve in very specific ways to form two and only two sets of large-scale \emph{chiral systems}. Each system contains spatially separated components with both signs of chirality, the upper portion having negative (positive) chirality and the lower part possessing positive (negative) chirality.  The components within a system are a filament channel (represented partially by sets of chromospheric fibrils), a filament (if present), a filament cavity, sometimes a sigmoid, and always an overlying arcade of coronal loops.  When we view these components as parts of large-scale chiral systems, we more clearly see that it is not the individual components of chiral systems that erupt but rather it is the approximate upper parts of an entire evolving chiral system that erupts.  We illustrate the typical pattern of build-up to eruptive solar events first without and then including the chirality in each stage of the build-up.  We argue that a complete chiral system has one sign of handedness above the filament spine and the opposite handedness in the barbs and filament channel below the filament spine. If the spine has handedness, the observations favor its having the handedness of the filament cavity and coronal loops above.  As the separate components of a chiral system form, we show that the system appears to maintain a balance of right-handed and left-handed features, thus preserving an initial near-zero net helicity.  We further argue that the chiral systems allow us to identify key sites of energy transformation and stored energy later dissipated in the form of concurrent CMEs, erupting filaments and solar flares.  Each individual chiral system may produce many successive eruptive events above a single filament channel.  Because major eruptive events apparently do not occur independent of, or outside of, these unique chiral systems, we hypothesize that the development of chiral systems: (1) are fundamental to the occurrence of eruptive solar events and (2) preserve an approximate balance between positive and negative helicity (right and left-handed chirality) while preparing to release energy in the form of CMEs, erupting filaments, and flares.  
\end{abstract}

\section{Introduction}
\subsection{Recognition of the Chiral Properties of Solar Features}
The concept of chiral systems emerged soon after the series of discoveries in the 1990s that many solar features possess handedness or \emph{chirality}.  We use the term \emph{chirality} in referring to specific observational signatures of handedness.  To discuss the related term \emph{helicity}, we must add either interpretation or measurements that respectively imply or apply mathematically definable properties.  Therefore, we begin with observations of the various forms of chirality characteristic of solar features and end with discussing the implications for understanding chirality as evidence of helicity.  In this conceptual paper we focus on the overall picture of the handedness of specific solar features, how they interrelate, and what we can learn from the overall picture.

Sunspot superpenumbral fibrils were the first solar features shown to have chirality by Hale (1927) and Richardson (1941). From studying rotational motions in chromospheric prominences, Gigolashvili (1978) concluded that there is a preference for left-hand spiral motions in prominences in northern solar hemisphere and the right-hand spiral motions for southern hemisphere.  The discovery of magnetic clouds in the interplanetary medium by Burlaga was followed by his recognition that these magnetic structures were either left-handed or right-handed magnetic flux tubes (Burlaga 1988).  Gosling (1990) schematically illustrated how the magnetic clouds could originate from sets of coronal loops   having footpoints in associated photospheric magnetic fields    skewed with respect to the polarity reversal boundary beneath the loops.  The coronal loops could be either left-skewed or right-skewed.     

The initial paper on the chirality of filament channels and the chirality of filaments (Martin, Bilimoria and Tracadas 1994) demonstrated the one-to-one relationship of dextral (sinistral) filaments barbs to the dextral (sinistral) fibril patterns of filament channels.  Their use of high quality H$\alpha$ large-scale images from the Big Bear Solar Observatory (BBSO) left no doubt that the respective chiralities of filament channels and filaments were determined by the direction of fibrils in the chromosphere and threads in the filaments.  The close and systematic relationship of filament threads to chromospheric fibrils provided strong confirmation that both fibrils and filament threads were field-aligned (Smith 1968).  This enabled the co-authors to deduce the overall rotational configuration of filament channels.  The rotational configuration can be considered as the total mean magnetic vector that a magnetometer would see if it was flown on a spacecraft from the positive network side of a filament, through the filament spine, and to the negative network side of the same filament. Without performing this experiment, the direction of the field-aligned fibrils of filament channels tell us unambiguously whether the magnetic vector rotates from vertically outward to the right or to the left to become coincident and parallel with the spine (Foukal 1971).  As redepicted in Figure 1, the fibrils on the positive field side of the filament are like arrows directly revealing the direction of the local magnetic field and indirectly indicating the magnetic field direction along the polarity reversal boundary (dashed line coincident with the filament) and in the field-aligned threads of a filament spine.

For dextral channels, the rotation of the magnetic vector across the channel (and across its polarity reversal boundary) from the positive to the negative network side of the channel is clockwise (right handed) as in Figure 1.  For sinistral channels with sinistral filament barbs, the rotation of the magnetic field vector is counterclockwise (left-handed).  (In Figure 1, the dextral configuration illustrated is the mirror image of the sinistral configuration of this channel on the Sun.)  

The samples of H$\alpha$ images used in the study of Martin, Bilimoria and Tracadas (1994) were only from special, high resolution observing runs to acquire data sets representing all categories of filament channels and filaments from all parts of the solar disk sampled equally. The high quality of the images enabled them to also establish the hemispheric dominance of one sign of chirality per hemisphere of filament channels and filaments and to definitively determine that the hemispheric pattern had exceptions.  Hence they were able to conclude that the hemispheric preference was a statistical relationship that would necessarily have a different origin than the invariable one-to one relationships found in the chiralities of the solar features irrespective of hemisphere.  The results of that study catalyzed other investigations into chiral properties of solar features.  Rust and Martin (1994) introduced the term chirality when they offered evidence that the chirality of the superpenumbral fibrils around sunspots also fit into the one-to-one relationship of filament channels and filaments. Pevtsov, Canfield and Metcalf (1994) found that active regions as whole could be characterized according to their helicity.  Martin and McAllister (1996) collected statistics on the chirality of coronal loops and flare loops, establishing that they too have one-to-one relationships with filament channels and filaments.  

\begin{figure}
\center
\includegraphics[scale=0.4]{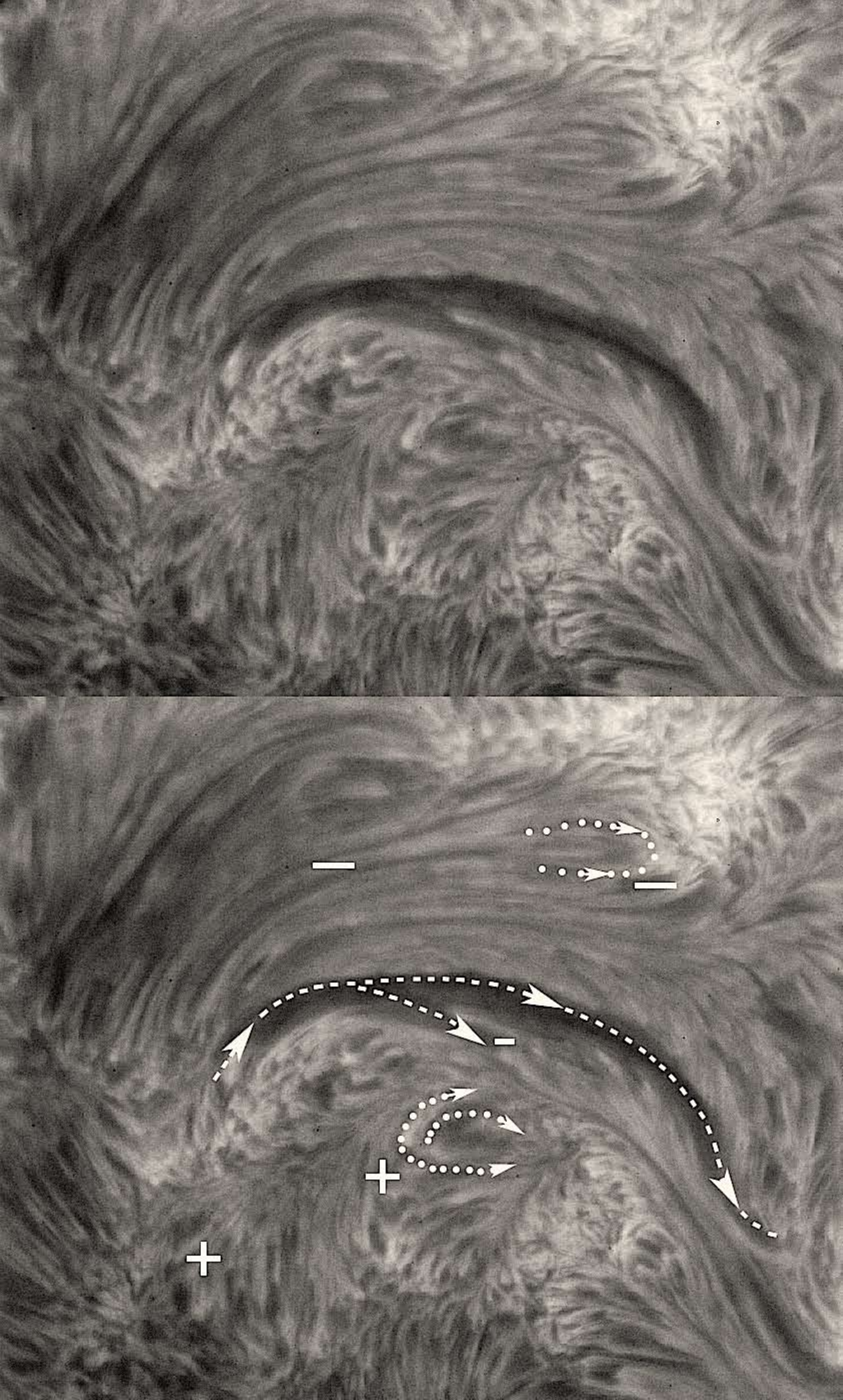}
\caption{The arrows and symbols on the lower image show how we interpret the direction of the local magnetic fields in fibrils in the chromospheric part of a filament channel and thereby deduce the direction of the magnetic field along the spine of the filament and at the footpoints of a barb.  The width of these images from the Dutch Open Telescope on 2010 Sep 28 is 113 arcsec.}
\end{figure}

Coronal loops and flare loops were found to be opposite in chirality to filament channels and filament barbs in the same system.  Because the chirality of all of these features are related to the polarity reversal boundary in the same way, one could not say that the opposing chiralities within a system  were due to any inconsistency in way the chirality of each of these features was defined.  The chirality of sigmoids (Rust and Kumar 1996) brought the second clear signature of chirality opposite to that of the filament barbs and filament channels.  Martin (1998a) combined all the forms of chirality into the pairs of diagrams in Figures 1 and 2 in that paper.  This made it clear that all solar features with chirality neatly fit into two and only two types of chiral systems that are opposites of each other.  The division, however, is not into systems that are completely right-handed or completely left handed.  Rather each system is split into neatly organized right-handed and left-handed components.  We further explore this idea in the present article.

Initially it was assumed by Martin, Bilimoria and Tracadas (1994), that the filament spine had no chirality because it could not be determined whether the ends of filaments terminated in the polarity boundary or in one polarity or the other.  Even at the highest spatial resolutions of current day observations (0.1 -0.3 arc sec), the question of whether filament ends terminate between polarities or in a given polarity has not been resolved observationally.  

If one pays attention to the shape of the spine, one can find clear cases that show the spine of some filaments have a slight reverse S-shape for filaments with dextral barbs and a slight S-shape for filaments that have sinistral barbs.  However, this is the case for only some filaments.  More filaments are slightly C-shaped (Martin 2003) like the one in Figure 1.  However, if one looks closely at the right end of the filament in Figure 1, it could be argued that as a whole the has a slight reverse-S-shape with the upper part of the reverse-S being very small, and lower part, large. This remaining uncertainty about the chirality of the spine is a negligible detail in the overall concept presented here.

The straight forward interpretation of filaments that are S-shaped or reverse S-shaped is that they are threads that run the full length of the spine, or long sections of the spine, and have the opposite chirality from the threads that form the barbs.  The finding by Wang, Muglach, and Kliem (2009) of displaced footpoints during the eruption of the spines of filaments also confirms that the spine is left handed (when its barbs are right-handed.  While the barbs cease their existence during or before eruption, the spine apparently can become left-handed due to magnetic reconnection with environmental fields during its eruption.  As pointed out by Ruzmaikin, Martin, and Hu (2003), this is not a paradox for right handed barbs to form from spine threads that are slightly left-handed if filaments are entirely composed of threads.  The spine and barbs can in principle have opposite chirality and helicity if they are different bundles of threads.  The different bundling of threads into these two categories is an important clue that the barbs must be formed by a different process than the threads of the spine. From studies of the filaments at highest sub-arcsecond resolution, this possibility is fortified by Lin (2004) and Lin et al. (2005) who have concluded that filaments are composed of thin, field-aligned threads.  In Figure 2 we illustrate the appearance of threads in a variety of filaments: (a) an active region filament, (b) an intermediate one (c) and (d) quiescent filaments.  These typical images, from time-lapse sequences recorded at the Dutch Open Telescope (DOT), all have dynamic threads that are seen as completely different in less than 5 minutes.

Confirmations that filaments are composed of field-aligned threads Lin, Martin and Engvold (2008) provides reasons to create schematic models of the 3D configurations of the magnetic fields of filaments as illustrated in Martin and Echols (1994), Lin, Martin and Engvold (2008) and Martin et al (2008).  In this paper we similarly use threads to represent filament spines and barbs but also use threads to represent all of the chiral features within a chiral system as shown in Figure 3.

\begin{figure}
\center
\includegraphics[scale=0.75]{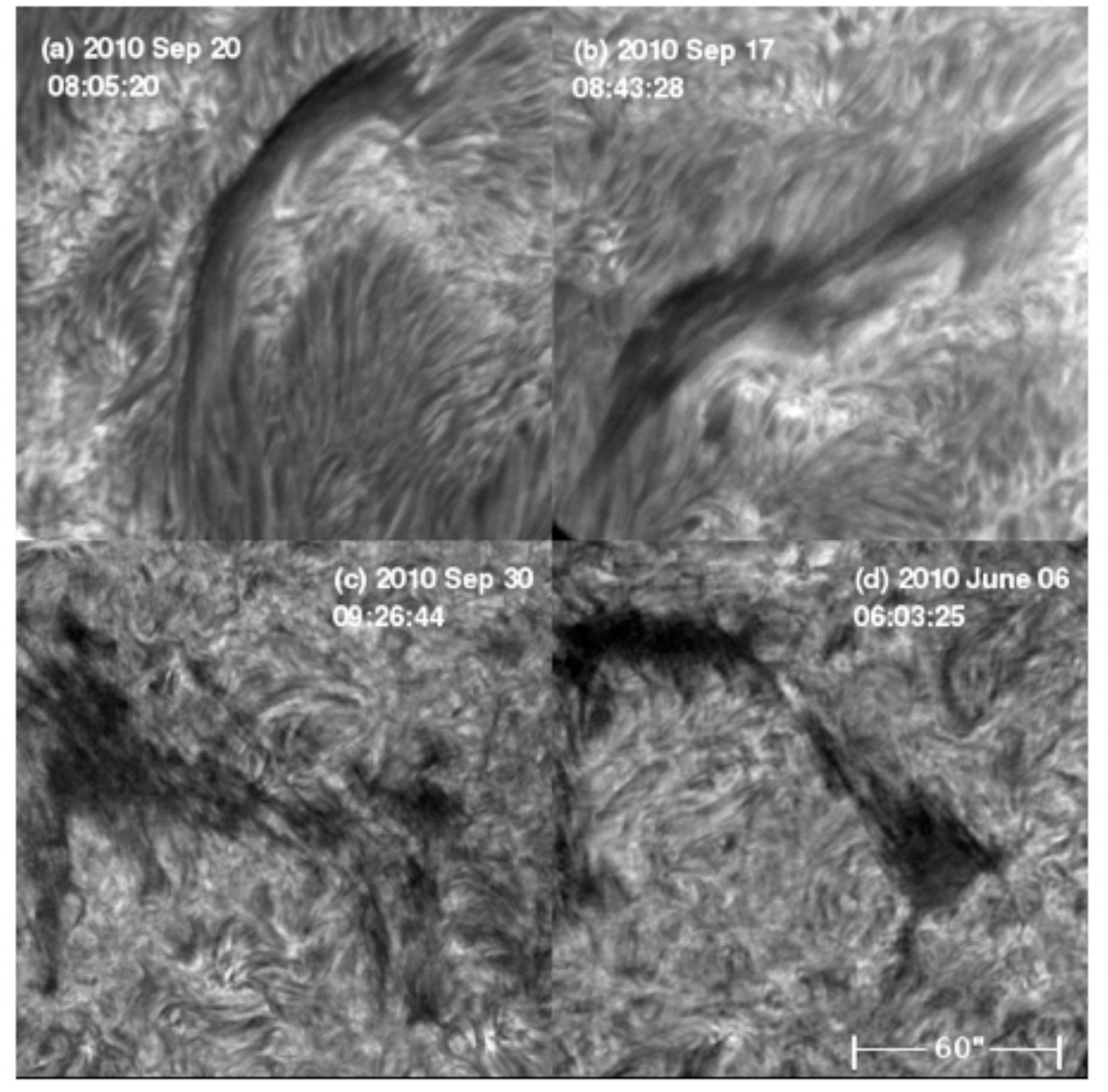}
\caption{The basic structure of filaments are fine threads whether in an active region with high magnetic flux density (a), in decayed active regions with intermediate flux density (b), or in quiescent filaments on the quiet Sun with low magnetic flux density (c and d). Some threads lie only along the spine (long axis) while others called \emph{barbs} deviate from the spine and connect to the adjacent chromosphere.  With lower magnetic flux density, the spines become high and less visible in H$\alpha$ but are usually more visible at the higher temperature 304 \AA line.  These H$\alpha$ images were recorded at the Dutch Open Telescope.}
\end{figure}

\subsection{Components of a Chiral System}
The key features of a chiral system are presented in Figure 3 giving four different views to show the three-dimensional character of each component.  Each of the features and its distinct chirality are color-coded as given in the right hand column.  The key components are:  (1) a coronal loop system (blue), (2) a filament channel with representative fibrils (green) (3) a filament with a spine (red), and (4) barbs (green),  and (5) a cavity, a region where the density at all wavelengths is extremely low all around the filament and within the coronal loop system.  These are the longest-lived of the features that exhibit chirality.  Such a chiral system is complimented on occasion by short-lived features that also have chirality, such as sigmoids or flare loops.  The chiralities of such short-lived features blend with the system as a whole.  

What properties of this combination of features shown in Figure 3 reveal it to be a chiral system?  The answer lies in the spatial organization of the chiral components into a simple but well ordered magnetic field system.  We answer the question by discussing the chiralities of features from the bottom to the top of the system.  

\begin{figure}
\center
\includegraphics[scale=0.3]{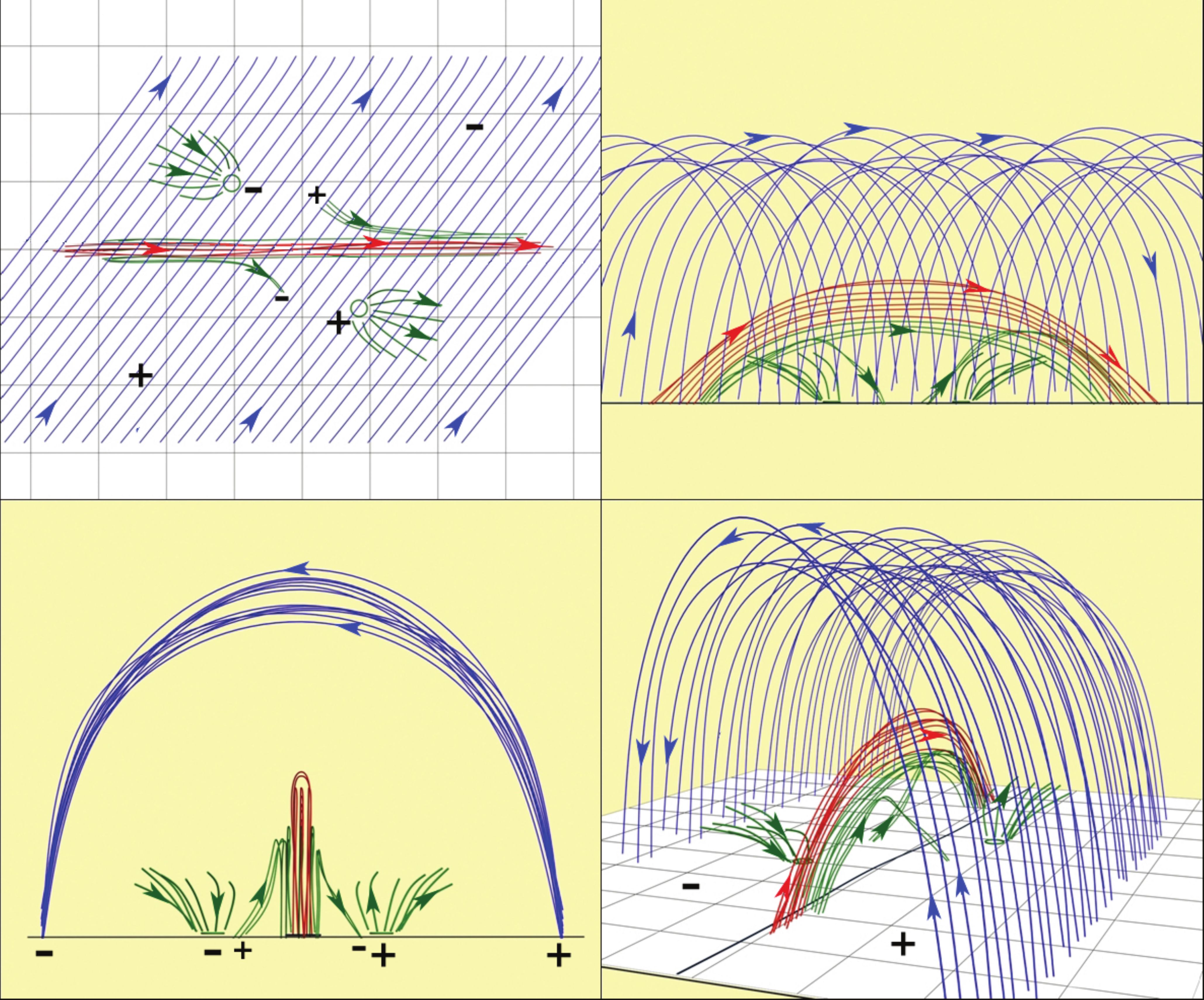}
\caption{The main components of a chiral system are the same features represented in the long term build-up to eruptive solar events except for cancelling magnetic fields which do not reveal chirality but are nevertheless fundamental to creating and maintaining a chiral system.  Cancelling magnetic fields result in the strong field component along the polarity reversal boundary in the photosphere above which we find the filament spine. The chiral system with five components: coronal loops (blue), chromospheric fibrils (green), filament spine (red), filament barbs (green), filament cavity: space between filament and overlying coronal loops.}
\end{figure}

\subsubsection{Filament channel}  
The filament channel is represented by the sets of green fibrils and their pattern is right-handed or dextral with respect to the polarity reversal boundary in the same pattern as shown for the filament channel in Figure 1.  Such a right-handed filament channel, along the polarity boundary, the field will point to the right as depicted in Figure 3.  

\subsubsection{Filament barbs}
Next to the fibrils near the polarity reversal boundary are the filament barbs.  The barbs do not join the chromospheric fibrils although they share a field component in the same direction along the channel.  In Figure 3, the barbs threads are right-handed or dextral.  They appear aligned with the nearest fibril in the top view but have an angle of about 90 degrees with respect to the fibrils as seen in the side and perspective views.  

\subsubsection{Filament spine}
At the top of the barbs is the filament spine that runs almost parallel with the polarity reversal boundary.  Physically, the spine can be viewed in either of two ways: (1) as a current sheet with its ends in the polarity boundary; in this case, it has no chirality or one might say it has zero helicity, (2) as long threads rooted in the photosphere as in Figure 3; in this case, the directions of the field along the spine threads is determined by the filament channel as discussed initially by Foukal (1971) and depicted in Figure 1.  For consistency with the field direction determined from the filament channel, and the schematic in Figure 3, the ends of the filament spine are shown as coming from the positive magnetic field side of the polarity boundary and ending in the negative side of the polarity boundary.  In this depiction, the filament spine is slightly left-handed.  

\subsubsection{Sigmoids}
When sigmoids are observed in the cavity of the system depicted in Figure 3, they have a distinct reverse S-shape and are therefore left-handed.  From this we infer that the cavity magnetic field is also left-handed (for the system shown in Figure 3).

\subsubsection{Coronal loops}
The coronal loop system at the top of the system is left-skewed in Figure 3 and is therefore left-handed relative to the polarity reversal boundary.   
\\
\subsection{Consistencies in the Overall Patterns of Chirality}
Looking at the organization of the features in Figure 3, we see that, with respect to the polarity reversal boundary, the spatial pattern is right-handed in the chromosphere and in barbs up to the height of the spine.  The spine is slightly left-handed and the cavity magnetic field, occasionally rendered partially visible by sigmoidal structures, is more strongly left-handed, that is, it has a larger angle with respect to the polarity reversal boundary.  The overlying loop system displays the largest angle with respect to the polarity reversal boundary.  Thus there is a rotation in the direction of the local magnetic field from bottom to top in the left-hand sense of 90 degrees or less.  However, in the direction across the filament channel at the height of the filament, the network magnetic field reveals an overall rotational pattern in the right hand-sense as schematically shown in Figures 1 in this paper and in Figure 3 in Martin, Bilimoria and Tracadas (1994).The right handed rotation of the inferred average magnetic field vector is approximately 180 degrees.  

It appears that there is a harmony of patterns among all of the chiral features.  In our schematic example, the approximate upper half of the system from the spine upward is left handed but below the spine the features are right-handed relative to the polarity reversal boundary.  For the chiral features as a group, there are approximately equal amounts of right-handedness and left-handedness.  This invariable systematic ordering of chiral features is the justifications for describing the set of chiral features as a Òchiral systemÓ.  The above example described one of two systems (i.e., left-handed above spine and right-handed below spine).  It is not necessary to describe the other system (i.e., right-handed above spine and left-handed below it) because all structures are the same except the signs of handedness for each feature are reversed from those presented in Figure 3. 

It is not accidental that the key long-lived features of this chiral system are the same features that are also the key players in the long-term build up to eruptive solar events.  Martin et al (2008) suggest that the formation of filament channels and filaments is the early part of a longer term build-up.  The long-term build-up begins with the formation of a filament channel and continues until the occurrence of a CME or more generally, an eruptive solar event usually including the eruption of a filament, a solar flare, and a CME.  In the present paper we take the concept of the long-term build-up to CMEs one step further.  We first summarize the stages in the build-up to eruptive solar events independent of their chirality as depicted under ÒKey ProcessesÓ in Figure 4.  Then, in Section 3, we present the build up to eruptive solar events as being synonymous with the development of chiral systems.  From this perspective, we can see   a chiral systems as a whole is more likely to be responsible for an eruptive solar event changes in an individual solar feature.  We emphasize that the purpose of this article is to provide a description of our concept.  How chiral systems might be modeled remains an open question.

\begin{figure}
\center
\includegraphics[scale=0.25]{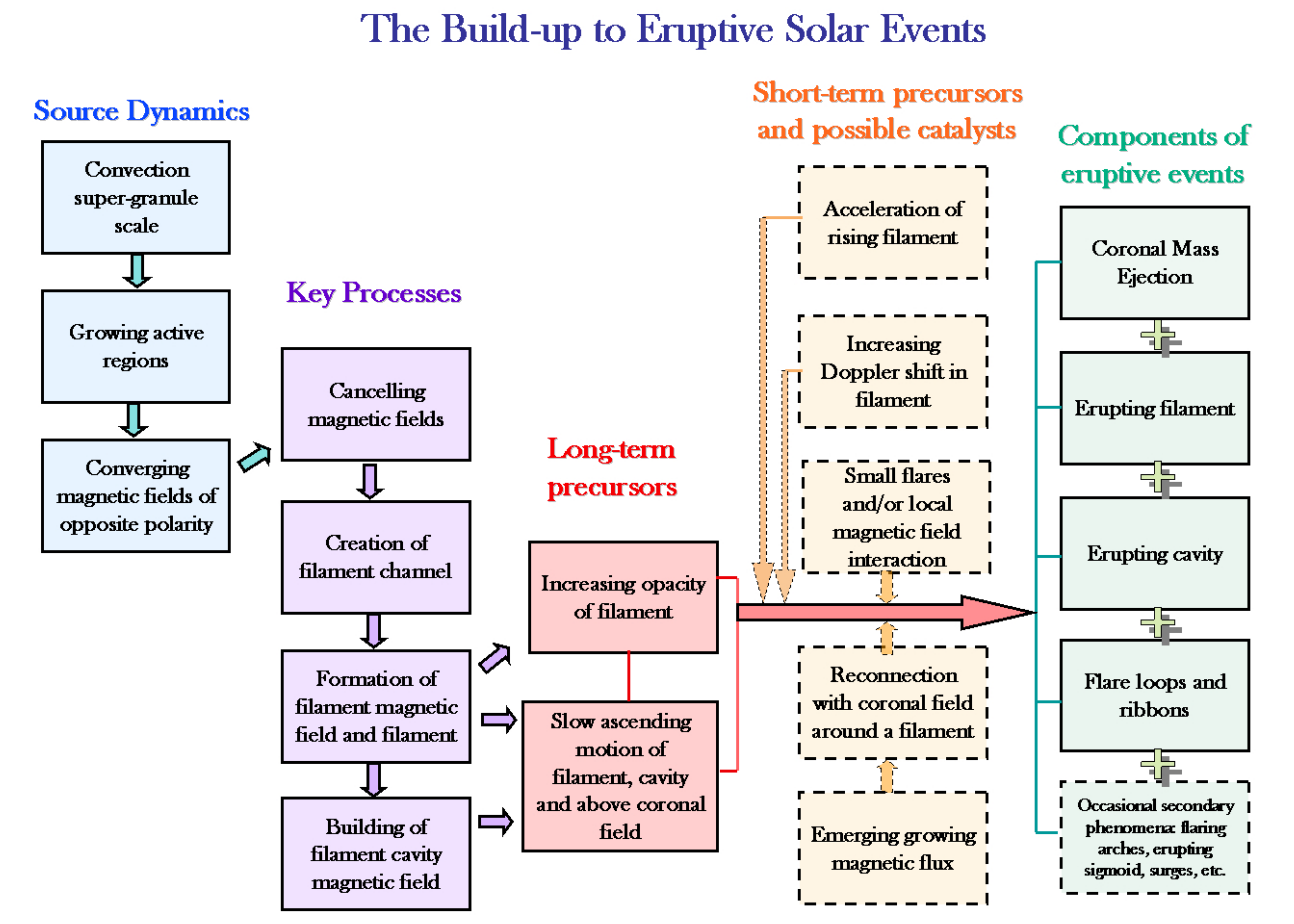}
\caption{Schematic representation of stages in the build-up to CMEs from the source dynamics and key processes in columns 1 and 2 respectively  to the components of eruptive solar events in column 5.}
\end{figure}

\section{The Long-term Build-up to CMEs }
The stages of the longÐterm build-up are shown in Figure 4.  The outcome of the build-up is an eruptive solar event consisting of a CME, an erupting filament and a flare.  The eruptive solar event is represented by the column of blocks on the extreme right side of Figure 4.  Placing the CME, erupting filament and flare in the same column is to indicate that the build up to CMEs is synonymous with the build up to eruptive solar events.  In our concept the buildup consists of all three components even when the erupting filament and flare are not necessarily visible or obvious.  

The essential linked stages in Figure 4 begin after the initial conditions in the first column giving the source of magnetic flux as being new active regions appearing at the solar surface and interacting with convection.  The build up in our picture is traced through a series of ensuing key processes shown in the second column.  Some of these steps are synonymous with the formation of filament channels and filaments (Martin 1990; 1998b).  The whole progression proceeds as follows:
    
\begin{enumerate}
\item Formation of coronal loop system between adjacent active regions, decaying active regions or the decayed remnants of one or more active regions     
\item Convergence and cancellation of magnetic fields along polarity reversal boundaries beneath the coronal loop systems which lead to
\item  Creation and maintenance of a filament channel related to the 
\item  Formation and evolution of filament threads which are a key to the 
\item  Building and growth of a filament cavity magnetic field leading sometimes to
\item Development of a sigmoid; however, with or without sigmoids, the building cavity magnetic field leads always to 
\item  The eruptive solar event consisting of a CME, often with an erupting filament, and a flare.
\end{enumerate}

Because our goal in this paper is a presentation of a broad concept, we do not attempt here to provide evidence of the validity of each step that we propose in our picture of the long-term build-up to eruptive solar events.  However, some of the basic picture is given in Martin et al. (2008).  We continue by focusing attention on the chirality of solar features in each stage of the build-up as schematically represented in Figure 5.

\section{The Build-up of a Chiral System}
The first stage in the development of a new coronal loop system is represented by the first row in Figure 5.  The coronal loop system could have formed between the opposite-polarity magnetic fields of two adjacent active regions.  Alternately, it could be the coronal loop system over a bipolar active region after new flux has ceased developing.  If the latter, the region is then in its initial stage of decay and both polarities of magnetic flux are beginning to spread in all directions from the local concentrations of magnetic flux of both polarities.  

Via these approximately random motions, some magnetic flux migrates toward the opposite polarity in the middle of the region while other flux encounters the opposite polarity around some of the boundary of the active region.  In either case, when the encounter of opposite polarity active region or network fields occurs, cancellation follows and the cancellation then results in the transformation of line-of-sight magnetic flux into a field component along the boundaries where the fields are cancelling in the manner described theoretically by Litvinenko, Chae and Park (2007) and this component results in the building of a filament channel.  This second stage of cancelling magnetic fields is not shown here; good examples are described in many papers such as those of Martin, Wang and Livi1985, Martin 1992 and Wang and Muglach (2007).

Filament channel formation is depicted in the second row in Figure 5; the fibrils from the plagettes (little plages above small network magnetic field clusters) have developed the characteristic pattern of a dextral filament channel.  On the positive side of the polarity reversal boundary, the magnetic field is directed from the plagette at the chromosphere outward along the spicules and into the corona.  On the negative side of the filament channel, the magnetic field goes into the Sun from the thin ends of the spicules down to the plagettes at the base of the spicules.  The development of this pattern is accompanied by the development of a strong horizontal magnetic field along the polarity reversal boundary in the photosphere, the chromosphere, and the low corona as illustrated by Martin and Panasenco (2010) and theoretically demonstrated by Litvinenko (2010).  Under a left-skewed coronal loop system, the fibril pattern changes from being similar in direction to the overlying coronal loop system to being up to 90 degrees with respect it.  This change is evidence that the filament channel is gradually storing magnetic field energy.  After the filament channel or a segment of a filament channel is completely formed, its configuration is the same as shown by Foukal (1971).  ÒCompletely formedÓ means that the magnetic shear has reached the maximum possible (Moore, Falconer and Sterling (2012) and that no fibrils cross the polarity reversal boundary where maximum shear exists (Panasenco 2010).  

The filament formation stage is represented in the third row in Figure 5.  This stage typically begins when the channel is completely formed (Martin 1998b).  However, this usually does not happen simultaneously all along the polarity reversal boundary.  Hence a filament often gradually lengthens as segments of the filament channel  arrive at the condition of maximum channel formation and maximum magnetic shear.  In our concept, formation of the filament magnetic field, represented in the third row of Figure 5, consists of two different processes:  (1) the formation of the spine and (2) the formation of barbs.  The condition for the formation of the filament spine is that cancelling fields continue along the polarity reversal boundary between network or active region magnetic fields of opposite polarity.  The filament magnetic field forms because there is no further possible change in the filament channel close to the polarity reversal boundary.  Therefore the energy of magnetic reconnection associated with the channel formation is literally channeled instead into increasing the magnetic field above the polarity reversal boundary. 

Whether mass input accompanies this stage of development of the spine or whether it occurs subsequently, or both, are still open questions.  The point is that the development of the filament magnetic field is considered here to be a continuation of the process that had been happening during the building of the filament channel.  In essence, the cancelling fields create new long magnetic threads from shorter ones.  As the reconnection sites are at the photosphere or very low in the chromosphere, it is reasonable that mass will be injected at the sites of reconnection.  This process has been best described to date by Litvinenko 1999, 1999 and Litvinenko, Chae and Park (2007).  However, if the rate of cancellation along the polarity reversal boundary is low, the channel will not be occupied by visible filament mass and is called an ÒemptyÓ filament channel (Pevtsov, Panasenco and Martin 2012).  The absence of visible mass in H$\alpha$ does not imply the absence of a filament magnetic field.
 
As depicted in the third row in Figure 5, the threads of filament barbs have one end merging with the spine and the other end connecting to the chromosphere adjacent to the spine near small areas that are opposite in polarity to the network on each side of a filament.  These small areas are called the ÒminorityÓ polarity; most of these fields originate within the centers of supergranules but some minor polarity patches are one pole of the numerous ephemeral active region magnetic fields.  It is rare for the strong-field network magnetic fields of opposite polarity to intermingle because they are confined to the boundaries of supergranules and can rarely migrate past one another without the complete cancellation of one polarity.   
The encounters of network of opposite polarities define the polarity reversal boundary which continuously shifts small amounts as patches of opposite polarity migrate together and cancel.    

\begin{figure}
\center
\includegraphics[scale=0.30]{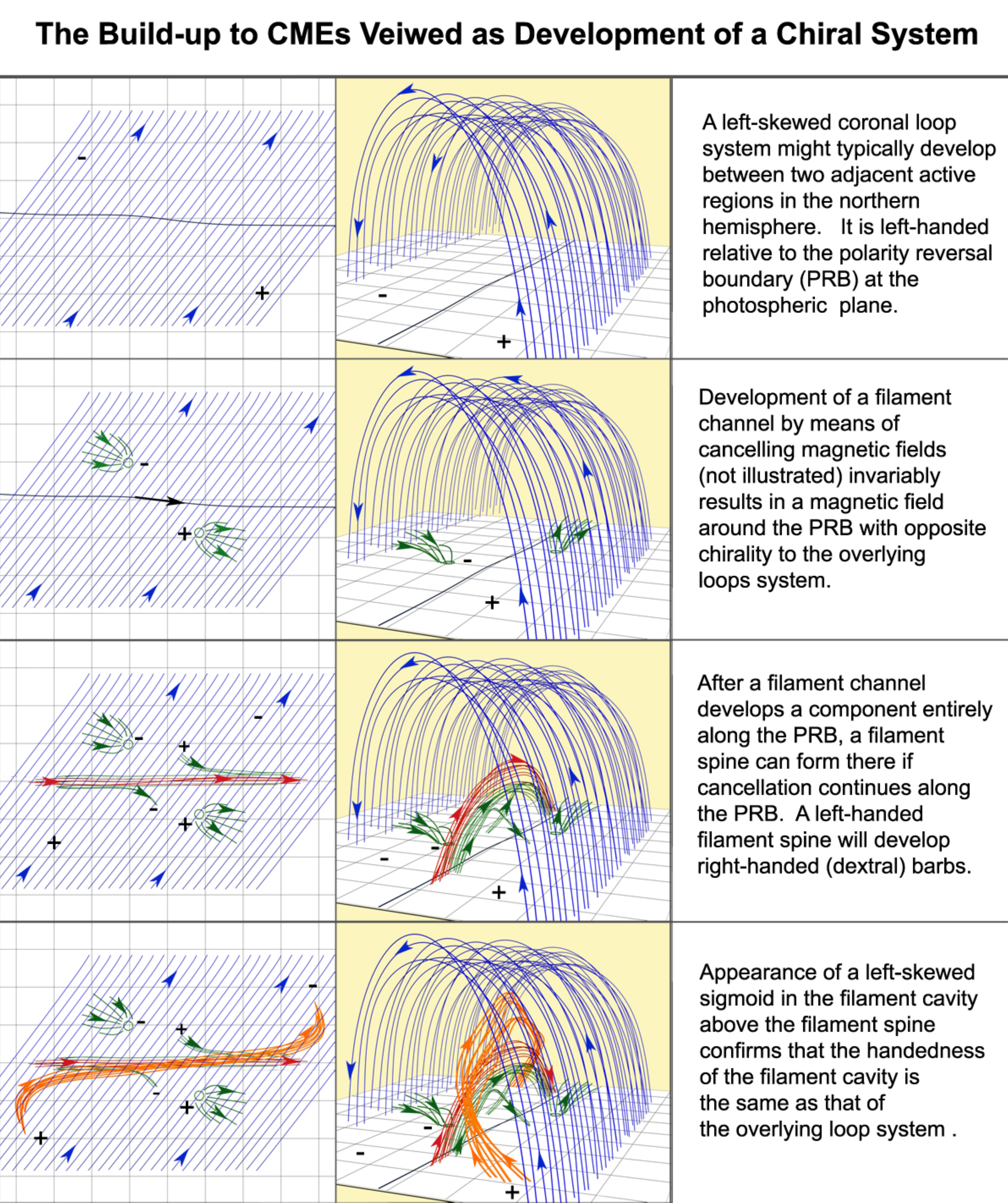}
\caption{A chiral system develops from the outside to the inside.  The initial coronal loops (left-handed in this example, are quickly balanced by the development of a right-handed filament channel.  In an intermediate filament channel, the development of a filament with a left-handed filament spine is accompanied by right-handed (dextral) barbs.  The right-handed barbs however, could be approximately balanced  by the presence of a left-handed filament cavity which is rendered visible by occasional sigmoidal structure.  These observations lead to our suggestion that the net helicity of a complete chiral system might be close to zero and that the conservation of helicity might apply to a chiral system or groups of chiral systems as a whole.}
\end{figure}

For barbs to have their observed characteristics, it is easily imagined that threads of the spine have undergone reconnection with an adjacent magnetic field, such as the ubiquitous intranetwork fields that are confined to supergranules or to the magnetic field of one pole of an ephemeral active region.  Such interactions would connect spine threads to the photosphere and should take place continuously but most would be too rapid and small scale to detect.  Only the magnetic fields of those threads aligned with the magnetic fields of the filament channel would survive long enough to have visible mass flows.  Therefore barbs represent only those residual products of magnetic reconnection between the spine and neighboring field that are parallel with the background magnetic field of the filament channel.  Therefore, the barbs give us essential information about the filament environment.  In the thread model, they do not need to share the chirality of the spine as they have been reconfigured by the local environment.  For this reason it is not surprising but rather necessary that barbs have the same chirality as the filament channel.  They are shown as dextral in Figure 5.

As inferred from Figure 5, the barbs provide the so-called inverse polarity of filaments as explained in Martin, Bilimoria, and Tracadas (1994).  Inverse meaning that their component of magnetic field perpendicular to the polarity reversal boundary is opposite in direction to the same perpendicular component of the coronal field lines that overarch filaments high above the spine.  One can then see that it is this opposite field direction that gives barbs the sign of chirality opposite to the overlying coronal loop system.  The barbs are merely manifesting the chirality of their filament channel. 
 
A final visible chiral feature shown in the fourth row in Figure 5 is a sigmoid.  Sigmoids appear in the filament cavity well above the filament spine and are either S or inverse S-shaped structures usually only seen in X-rays high temperature lines.  For reasons not yet understood, some specific local fields within the cavity become filled with mass and radiate in X-rays and at EUV wavelengths.  For those sigmoids that are situated above filaments, their chirality is always the same sign as the coronal loops in their chiral system (Martin 1998a).  From the strong tendency of magnetic fields to remain smooth, we infer from the sigmoids that the entire cavity has the sign of helicity of the overlying coronal loops system and is intermediate in its degree of chirality between the filament spine (which is either zero or minor) and the strong skew of the overlying loop system.

\section{A Final Overview:  The Significance of Chiral Systems}

\subsection{The link of CMEs to filament channels}
Because CMEs only occur above filament channels (Martin et al 2008), there is an inevitable link between CMEs and the development of chiral systems that include filament channels and filaments.  In this paper, we have reviewed some of the basic properties of chiral systems and have shown that components of these systems are also the components that play major roles in the build-up to eruptive solar events.  In our view the significance of the chiral properties of solar features is that they together reveal the existence of chiral systems. Recognition of chiral systems allows us to better appreciate that eruptive solar events are not driven by a single physical process but rather by a series of interrelated processes.  

\subsection{A balance of left and right handed features}
From our description of the build-up of chiral systems, it follows that the approximate upper parts of the system have one sign of chirality while the lower parts have the opposite signs of chirality.  The division into left-handed and right-handed domains occurs close to the height of the spine.  Our conception of the process of barb thread formation from spine threads causes them to acquire and maintain the chirality of the lower half of a system.  This is possible in the empirical thread model of filaments (Lin, Martin, and Engvold 2008) that has been presented as an observational guide towards the development of future quantitative models of filaments.

\subsection{Assumptions about cause and effect between components\\ within a chiral system}
The recognition of chiral systems broadens our understanding so that we see that every component or active process within a chiral system is an indicator of what is happening or building in that part of the system rather than evidence that one component triggers another.  In recent years, we have witnessed the frequent reiteration of the broader understanding that flares, erupting filaments and CME do not cause one another; rather they all have a common cause.  Similarly, the individual components of chiral systems do not cause one another.  Rather each contributes or plays a role in revealing the overall behavior of the system as a whole.  From this perspective, we can advance to a more complete understanding of eruptive solar events and to the possibilities that the behavior of such systems as a whole might also become predictable.  

\subsection{From chirality to helicity}
From the twist of both sunspot superpenumbral fibrils and rotational magnetic fields of magnetic clouds, it was readily recognized that the handedness of the magnetic fields of these features represented helicity.  Helicity can be realized in several forms (Berger 1998, 1999). For a single flux rope, as interpreted for explaining magnetic clouds in the interplanetary medium, their helicity decomposes into contributions from field lines twisting about the axis (twist $T$) and the coiling of the axis itself (writhe $W$): 
\begin{equation}
H = 
\left(T+W\right)\Phi^2,
\end{equation}

where $\Phi$ is the total magnetic flux (Berger and Prior 2006). For a collection of magnetic features, the helicity decomposition is more complicated. Helicity is then expressed as a sum over self-helicities of individual components (e.g. twist and writhe) and mutual helicities between components (e.g. shear and linking).

Recently, studies of both theory and observations have investigated whether magnetic helicity buildup in the corona is associated with magnetic activity (Pevtsov 2008). Helicity flux through the photosphere can be measured using the motion of magnetic flux elements (Pariat et al 2005). Observed helicity flux into active regions correlates with increased flaring activity (Chae 2007, Park et al 2010). While some of the helicity flux may arise from differential rotation of pre-existing flux, there are indications that additional helicity must be injected, possibly from below the photosphere (Pevtsov et al 2003).  

The development of filament channels at the photosphere and above is evidence for another physical process that results in the buildup of helicity at the solar surface.  The buildup of filament channels is linked observationally to cancelling magnetic fields which, in turn, has been attributed to magnetic reconnection at or very close to the photosphere (Litvinenko, 1999; Litvinenko, Chae, and Park (2007). We suggest that this mechanism for the cancelling magnetic fields can be a primary source of the buildup of energy and helicity in active regions, specifically at the sites of filament channels and possibly at any location where cancellation is observed.  This means of helicity buildup can be important from the very beginning of active regions because cancellation takes place between elementary bipoles in the earliest stages of active region formation (Martin 1990).

Why do solar eruptive events occur? The initial broad answer has been because the Sun must expel energy built up in the solar atmosphere. However, in recent years, there has been a different new broad answer.  The Sun must shed helicity that builds up in the solar atmosphere. This idea was first presented by Rust (1999) and by Low (1995, 2001). A study of Georgoulis et al (2009) indicates that the total amount of helicity from photospheric magnetic fields matches very closely the amount of helicity expelled by the coronal mass ejections over the solar cycle 23. In other words, there is no accumulation of net helicity on the Sun. We suggest that the more complete answer is to shed energy and helicity together.

This paper has presented a broad picture of the interplay between steps in the energy buildup and the helicity buildup in a combination of features that we define as chiral systems. These erupting chiral systems do shed helicity from the Sun concurrent with many forms of significant energy release. By defining the components and illustrating properties of chiral systems, a greater grandeur unfolds. The concept of chiral systems are at the same time both more complex in detail but intriguing in its overall simplicity; Chiral systems consist of two and only two patterns that are mirror images of each other. 

In Figure 4, we note that the left-handed coronal loops and right handed filament channels most likely develop simultaneously. This simultaneous development might help in maintaining a balance between negative and positive helicity respectively throughout the building of each chiral system, maintaining the conservation of total helicity.  In the majority of cases in the northern hemisphere, the overlying arcade is skewed in the left hand sense in the northern hemisphere, while the filament channel below it concurrently balances this negative chirality with the building of positive chirality between the footpoints of coronal arcade in the chromosphere and photosphere; exceptions to the hemispheric pattern are just the opposite;  in the same hemisphere, some chiral systems develop that are mirror patterns  of the majority pattern influenced by differential rotation. These exceptions demonstrate that the root cause of chiral systems cannot be differential rotation alone although differential rotation plays an important role in the overall global picture.

The divisions of solar chiral systems into positive and negative components allow us to see the possible application of the conservation of helicity on a local scale in addition to the global scale such as that generated by the differential rotation of the Sun.  The local helicity as seen in chiralities of solar features develops much too rapidly to be accounted for by solar rotation or to be balanced only by the opposite helicity in the opposite hemisphere. An example is in the rise of current solar cycle 24.  The northern hemisphere is arriving at its maximum months to years ahead of the southern hemisphere.  If the global balance of helicity involves temporary helicity storage in deep subsurface fields, we should then expect that coronal helicity balance might occur within each chiral system or group of chiral systems.  
    
In the system illustrated, we note that the building filament spine appears to retain negative chirality to the system while the creation of barbs removes a part of this negative chirality and replaces it with positive chirality.  In the short-term, we suggest that the helicity of the building chiral system is approximately conserved.  We also infer that the spatial scale of the chiral systems increases with time as magnetic energy accumulates within the system.  This then forms the basis of a new hypothesis that chiral systems preserve a balance between positive and negative helicity while preparing for the release of their accumulated energy in the form of CMEs, erupting filaments and flares. 

To test this concept, we suggest that attempts could be made to calculate the total helicity of a chiral system, necessarily including mutual helicity only briefly mentioned here. We suggest that the simplest isolated cases for testing this concept are those in which an active region and at least one subsequent chiral system develop in a coronal hole in one of the two main latitude belt of solar activity.  We further suggest that efforts to quantify the development of chiral systems will eventually lead to a more complete understanding of the interplay between helicity and the storage of magnetic energy in eruptive solar events. 

\acknowledgments
SFM and OP acknowledge NSF grant AGS 0837915 and NASA grant NNX09AG27G. AAP acknowledges support from the National Solar Observatory (NSO) operated by the Association of Universities for Research in Astronomy, AURA Inc under cooperative agreement with the National Science Foundation (NSF). Work by NS partially contributes to the research for
European Union Seventh Framework Programme (FP7/2007-2013) for the
COronal Mass Ejections and Solar Energetic Particles(COMESEP) project
under grant agreement 263252. The images in Figures 1 and 2 were obtained at the Dutch Open Telescope (DOT) under NSF Rapid Award AGS-1024793 to the Helio Research team including co-authors OP, OE, YL, NS, AAP, and SFM who thank R.H. Hammerschlag and G. Sliepen for collaborative observations throughout the 2010 observing season. The Technology Foundation STW in the Netherlands financially supported the development and construction of the DOT and follow-up technical developments. We thank V. Gaizauskas for making available the H$\alpha$ Lyot filter at the DOT.  

\section{References}

Berger, M.A. 1998, Magnetic Helicity and Filaments. In: Webb, D., Rust, D. and Schmieder, B. (eds) New Perspectives on Solar Prominences. IAU Colloquium 167, Ast. Soc. Pac., San Fransisco, 102-110.

Berger, M.A. 1999, Introduction to magnetic helicity, in Plasma Phys. Controlled Fusion, Vol. 41, p. B167 Ð B175.

Berger,  M.A.; Prior, C., The writhe of open and closed curves, J Phys A-Math Gen, volume 39:8321-8348.

Burlaga, L. F. 1988, JGR, 93, 7217 

Chae, J. 2007 Measurements of magnetic helicity injected through the solar photosphere Advances in Space Research 39, 1700-1705

Park, Sung-hong, Chae, Jongchul, and Wang, Haimin 2010,Productivity of Solar Flares and Magnetic Helicity Injection in Active Regions,  Ap. J. 718:43Ð51

Foukal, P. 1971, Morphological Relationships in the Chromospheric H$\alpha$ Fine Structure, Solar Phys. 19, 59  

Gosling, J. T. 1990, Coronal mass ejections and magnetic flux ropes in interplanetary space, in Physics of magnetic flux ropes, Washington, DC, American   
                       Geophysical Union, 1990, p. 343-364.

Georgoulis, M.K., Rust, D.M., Pevtsov, A.A., Bernasconi, P.N., Kuzanyan, K.M.: 2009,
"Solar Magnetic Helicity Injected Into the Heliosphere: Magnitude, Balance, and Periodicities over Solar Cycle 23", Astrophys. Journal, 705, L48-52

Gigolashvili, M. Sh. 1978, An Investigation of Macroscopic Motions Using the Ca+ Lines in the Prominence of 15 October 1969, Solar Phys. 60, 293-298

Hale, G. E., 1927, Nature, 119, 708.

Lin, Y.:  2004, Magnetic field topology inferred from studies of fine threads in solar filaments, PhD Thesis, Institute of Theoretical Astrophysics, University of Oslo

Lin, Y., Martin, S.F. and Engvold, O. 2008, Filament substructures and their interrelation, eds. R. Howe, K.S. Balasubramaniam and R. Komm, and G.J.D Petrie, ASP Conference Series, Vol. 383, 235-242. 

Lin, Y.; Engvold, O.; Rouppe vanderVorrt,Luc; Wiik, J. E. and Berger, T.E. 2005, Thin Threads of Solar Filaments, Solar Phys. 226, 239-254.    
  
Litvinenko, Y. E. 1999, Photospheric Magnetic Reconnection and Canceling Magnetic Features
on the Sun, Astrophys. J. 515, p. 435-440.
  
Litvinenko, Yuri E. 2010, Evolution of the Axial Magnetic Field in Solar Filament Channels, Astrophys. J. 720, p. 948-952.
  
Litvinenko, Y., Chae, J., and Park, S.-Y.: Flux Pile-up Magnetic Reconnection in the Solar Photosphere, Astrophys. J. 662, p. 1302-1308 (2007)
  
Low, B.C. 1996, Solar Activity and the Corona (invited review), Solar Phys. 167, 217-265.  
  
Low, B.C.  2001, Coronal mass ejections, magnetic flux ropes, and solar magnetism, J. Geophys. Res. 106, A11, p.25141-25164.
    
Martin, S.F. and Livi, S.H.B. 1992, The role of canceling magnetic fields in the build-up to erupting filaments and flares, in Eruptive Solar Flares, eds. Z Svestka, B.V. Jackson, and M. Machado, Lecture Notes in Physics, Springer-Verlag, p. 33-45.  
    
Martin, S.F. 1998, Filament Chirality: A link between fine-scale and global patterns. In Webb, D., Rust, D.M., Schmieder, B. (eds.) New Perspectives on Solar Prominences. Kluwer Acad. Pub., Dordrecht, 419-429.
    
Martin, S. F. 1990, ÒElementary Bipoles of Active Regions and Ephemeral Active RegionsÓ Societa Astronomica Italiana, Memorie (ISSN 0037-8720), vol. 61, no. 2, 1990, p. 293-315.
    
Martin, S.F. 2003, ÒSigns of Helicity in Solar Prominences and Related FeaturesÓ Adv.  Space Res. 32, 1883-1893.
    
Martin, S.F. 1998a, Conditions for the Formation and Maintenance of Filaments (Invited Review), Solar Phys., 182, p. 107-137
    
Martin, S.F. 1998b, Filament Chirality: A Link between fine-scale and global patterns, in New Perspectives on Solar Prominences,  eds. Webb, D., Rust, D.M. and Schmieder, G.,  IAU Colloquium 167, ASP Conf. Series 150, 419-438
    
Martin, S. F.; Bilimoria, R.; Tracadas, P. W. 1994 Magnetic field configurations basic to filament channels and filaments. Solar Surface Magnetism. NATO Advanced Science Institutes (ASI) Series C: Mathematical and Physical Sciences, Proceedings of the NATO Advanced Research Workshop, held Soesterberg, the Netherlands, November 1-5, 1993, Edited by Robert J. Rutten and Carolus J. Schrijver. Dordrecht: Kluwer Academic Publishers, p.303.
    
Martin, S.F. and Echols, C.R. An observational and conceptual model of the magnetic field of a filament, in Solar Surface Magnetism, eds. R.J. Rutten and C.J. Schrijver, Kluwer Academic Publishers, Dordrecht, p. 339-346 (1994)
    
Martin, S. F., Lin, Y., Engvold, O.  2008, A Method of Resolving the 180 Degree Ambiguity Employing the Chirality of Solar Features. Solar Physics, 250, 31-51.
    
Martin, S.F., Panasenco, O.; Engvold, O. and Lin, Y. 2008, ÒThe Link Between CMEs, Filaments, and Filament ChannelsÓ Annales of Geophysicae 26, 3061-3066.
Martin, S.F. and Panasenco, O. 2010, Memorie della Societa Astronomica Italiana, vol. 81, p. 662.
    
Martin, S. F.; McAllister, A. H. 1996,The Skew of X-ray Coronal Loops Overlying H alpha Filaments, in Magnetodynamic phenomena in the solar atmosphere. Prototypes of stellar magnetic activity, Proceedings of IAU 153rd colloquium 153, Makuhari; near Tokyo; Japan; May 22-27; 1995; Dordrecht: Kluwer Academic Publishers,  ed. Yutaka Uchida, Takeo Kosugi, and Hugh S. Hudson, p. 497. 
    
Martin, S.F.,  Livi, S.H.B. and Wang, J., The Cancellation of Magnetic Flux. II - In a Decaying Active Region (R. G. Giovanelli Commemorative Colloquium, Tucson, AZ, Jan. 17, 18, 1985) Australian Journal of Physics (ISSN 0004-9506), vol. 38, 1985, p. 929-959 (1985)
    
Moore, R. L.; Falconer, D. A.;  Sterling, A. C.  2012, The Limit of Magnetic-Shear Energy in Solar Active Regions, accepted by Astrophys. J. Feb 2012.
    
Pariat, E., Demoulin, P. and Berger, M.A. 2005 Photospheric flux density of magnetic Helicity, Astronomy and Astrophysics 439 1191Ð12   
    
Pevtsov, A.A.: 2008, "What Helicity Can Tell us About Solar Magnetic Fields", 
J. Astrophys. Astron., 29, 49-56

Panasenco, O. 2010, Memorie della Societa Astronomica Italiana, 81, 673
    
Pevtsov, Alexei A.; Canfield, Richard C.; Metcalf, Thomas R. 1994, Patterns of helicity in solar active regions, Ap. J. 425, L117-L119.
    
Pevtsov, A. A., Maleev, V. M, and Longcope D. W. 2003, "Helicity Evolution in Emerging Active Regions", Astropphys. J. 593, 1217-1225. 
    
Pevtsov, A.A., Panasenco, O. and Martin, S.F.: 2012, "Coronal Mass Ejections from Magnetic Systems Encompassing Filament Channels without Filaments", Solar Physics, 277, 185-201
    
Richardson, R.S.  1941, The Nature of Solar Hydrogen Vortices, Ap. J. 93,  24-28.
    
Rust, D.M., Kumar, A. 1994, Helical magnetic fields in filaments. Solar Phys. 155, 69-97
    
Rust , D.M. Kumar, A. 1996, Astrophys. J. 464, L199.
    
Rust, D.M. and Martin, S.F. 1994, A Correlation Between Sunspot Whirls and Filament Type, in Solar Active Region Evolution: comparing Models with Observations, ASP Conf. Series, Vol. 68, 337.
    
Rust, D. M. 1999, Magnetic helicity in solar filaments and coronal mass ejections. Geophys.  Monogr. Ser. 111, AGU, 221-227.
    
Ruzmaikin, A.; Martin, S.F. and Hu, Q. 2003, ÒSigns of magnetic helicity in interplanetary coronal mass ejections and associated prominences: Case studyÓ J. Geophys. Res. 108, 1096.
    
Smith, Sara F. 1968, The Formation, Structure and Changes in Filaments in Active Regions, Structure and Development of Solar Active Regions. Symposium no. 35 held in Budapest, Hungary, 4-8 September 1967. Edited by Karl Otto Kiepenheuer. International Astronomical Union. Symposium no. 35, Dordrecht, D. Reidel., p.267
    
Wang, Y-M. and Muglach, 2007, K. On the Formation of Filament Channels, Astrophys. J., 666, p. 1284-1295 (2007)
    
Wang, Y.-M.; Muglach, K.; Kliem, B. 2009, Endpoint Brightenings in Erupting Filaments, Astrophysical J. 699, 133-142.

\end{document}